\documentclass[10pt,twocolumn,twoside] {IEEEtran}

\usepackage{cite}

\ifCLASSINFOpdf

\else
   \usepackage[dvips]{graphicx}
   \DeclareGraphicsExtensions{.eps}
\fi
\usepackage[cmex10]{amsmath}
\usepackage{algorithmic}
\usepackage{array}
\usepackage{amsthm}
\theoremstyle{remark}
\newtheorem{theorem}{Theorem}
\newtheorem{lemma}{Lemma}
\newtheorem{remark}{Remark}
\usepackage{booktabs}
\usepackage{bm}
\usepackage{tabularx}
\begin{document}
\title{Off-grid DOA Estimation Based on Analysis of the Convexity of Maximum Likelihood Function}
\author{Liang~Liu
        and~Ping~Wei
\thanks{The authors are with the School of Electronic Engineering, University of
Electronic Science and Technology of China, Chengdu 611731, China (e-mail: liu\_yinliang@outlook.com; pwei@uestc.edu.cn).}}

\markboth{Journal of \LaTeX\ Class Files,~Vol.~11, No.~4, December~2012}%
{Shell \MakeLowercase{\textit{et al.}}: Bare Demo of IEEEtran.cls for Journals}

\maketitle

\begin{abstract}
Spatial compressive sensing (SCS) has recently been applied to direction-of-arrival (DOA) estimation owing to advantages over conventional ones. However the performance of compressive sensing (CS)-based estimation methods decreases when true DOAs are not exactly on the discretized sampling grid. We solve the off-grid DOA estimation problem using the deterministic maximum likelihood (DML) estimation method. In this work, we analyze the convexity of the DML function in the vicinity of the global solution. Especially under the condition of large array, we search for an approximately convex range around the ture DOAs to guarantee the DML function convex. Based on the convexity of the DML function, we propose a computationally efficient algorithm framework for off-grid DOA estimation. Numerical experiments show that the rough convex range accords well with the exact convex range of the DML function with large array and demonstrate the superior performance of the proposed methods in terms of accuracy, robustness and speed.
\end{abstract}


\IEEEpeerreviewmaketitle

\section{Introduction}

\IEEEPARstart{T}{his} paper addresses the problem of off-grid direction-of-arrival (DOA) estimation through analysis of the convexity of the maximum likelihood (ML) function. The ML criterion has drawn much attention due to its attractive benefits \cite{Sandkuhler1987,Stoica1989,Stoica1990,Stoica1990a,Stoica1990b}, such as consistency, asymptotic normality, efficiency. However, the ML estimation approach generally requires a multidimensional search at a great computational cost in practical multi-target scenarios. Fortunately, some iterative algorithms, such as the Newton type algorithm\cite{Gill1981,DennisJr1996},  are applied to reduce computation cost. However, those iterative algorithms need a good initial value, which must be ¡°sufficiently close¡± to the global minimum in order to prevent convergence to a local extremum. This work is dedicated to find out a convergence region by analyzing the convexity of the ML function.

Array processing for DOA estimation has been a topic of intensive research interest during the past two decades\cite{Krim1996}. Since the ML approach is often deemed exceedingly complex, several suboptimal methods are proposed, such as subspace based methods, especially MUSIC algorithm\cite{Schmidt1986}, which exchange the multidimensional search problem for one-dimensional search problem. Whereas, the performance of MUSIC algorithm is not good as ML estimation in DOA estimation with snapshot deficient scenario or coherent signals\cite{Stoica1989,Stoica1990,Stoica1990a}.

Recently, the research on DOA estimation has been advanced owing to the development of methods based on compressed sensing (CS) or sparse signal reconstruction (SSR) \cite{Donoho2006} by exploiting the spatial sparsity in the array model. The CS-based DOA estimation approaches are extremely attractive for their ability to resolve closely spaced sources, few snapshots and correlated sources. Some CS-based DOA estimation methods have been presented. $\ell_{1}$ optimization (or called basis pursuit, BP) \cite{Chen1998,Candes2006,Li2008}, matching pursuit (MP) \cite{Mallat1993,DeVore1996}, and sparse Bayesian inference/learning (SBL) \cite{Tipping2001,Ji2008,Babacan2010} are proposed in the case of a single measurement vector (SMV). In the case of multiple measurement vectors (MMV), the simultaneous sparse approximation problem arises, and singular value decomposition (SVD) is introduced to reduce complexity and sensitivity against noise \cite{Malioutov2005}. A natural extension of basis pursuit is convex relaxation algorithms, which employ mixed norm $\ell_{p,q}$ optimization in this case \cite{Cotter2005,Eldar2009,Tropp2006,Hyder2009,Malioutov2005}. Simultaneous orthogonal matching pursuit (S-OMP) \cite{Cotter2005} or multiple response model orthogonal matching pursuit (M-OMP) \cite{Tropp2006a}, is a multiple response variant of matching pursuit. Additionally, sparse Bayesian learning for MMV guarantees \cite{Wipf2007} the simultaneous sparsity by assuming the same sparse prior, such as a Laplace signal prior, for the signals at all snapshots.

All these CS-based DOA estimation methods employ fixed sampling grid and assume that all the true DOAs are exactly located on the selected grid. When the true DOAs are beyond the fixed grid, their performance will degrade due to discretization error. There are still difficulties to set the grid interval in practical situations where the true DOAs are not on the sampling grid. On one hand, a dense sampling grid is necessary for accurate DOA estimation to reduce the gap between the true DOA and its nearest grid point since the estimated DOAs are constrained on the grid. But on the other, a dense sampling grid leads to high computational complexity of recovery algorithm and a highly coherent matrix that violates the condition for the sparse signal recovery.

In \cite{Zhu2011}, an off-grid model for DOA estimation is introduced and a sparse total least squares (STLS) method based on the Gaussian assumption of off-grid distance is proposed. However, the Gaussian condition cannot be satisfied in the off-grid DOA estimation problem. In \cite{Yang2013}, a new off-grid algorithm termed OGSBI-SVD from a Bayesian perspective is presented by assuming the off-grid distance satisfies uniform distribution. As shown in \cite{Yang2013}, OGSBI-SVD can exceed a lower bound of root mean squared error (RMSE) that is shared among all on-grid model based methods. But its performance is not ideal especially with high signal to noise ratio (SNR) or large snapshots. Besides, this approach is not robust in the case of large grid interval.

This work proposes a deterministic ML (DML) based approach to mitigate CS-based DOA estimation biases. This approach presents the same performance as DML, whose asymptotic estimation performance often able to achieve the Cram{\'{e}}r-Rao bound (CRB). So this approach is no longer limited by the density of sampling grid. This work can be viewed from two aspects, one is that it mitigates CS-based DOA estimation biases through ML estimation, the other is that it realizes the ML estimation through CS-based DOA estimation and Newton iteration. It is the key issue to make the approach convergent. The main contributions of this work are listed as follows. Firstly, we prove that the DML function for DOA estimation is a convex function in the vicinity of the global minimum. Secondly, we come to a conclusion that an approximate convex range of the likelihood function can be obtained under certain conditions, which to some extent suggests how ``sufficiently close'' to the global minimum the iterations must be initialized. Therefore, we decide the density of sampling grid for CS-based DOA estimation algorithms by our result instead of empiric value. In addition, we propose a class of algorithms for off-grid DOA estimation based on the convexity of the DML function. The proposed algorithms consist of two stages. In the first stage, by setting the grid interval according to our result, CS-based DOA estimation approaches obtain coarse DOA estimates. In the second, we initialize the Newton type iteration with these estimated DOAs to realize the DML estimation. Because of the convexity of the DML function, our proposed methods are likely to converge to the global minimum of the DML function.

The following notations are used in the paper.  ${\left(  \cdot  \right)^{\rm T}}$ and ${\left(  \cdot  \right)^{\rm H}}$  denote the transpose and Hermitian transpose, respectively. ${\left\|  \cdot  \right\|_F}$, $E\left(  \cdot  \right)$,  ${\mathop{\rm Tr}\nolimits} \left(  \cdot  \right)$ and $\Re \left(  \cdot  \right)$ stand for the Frobenius norm, expectation, trace, and real part operators, respectively. $diag\left( {\bf{A}} \right)$ denotes a column vector composed of the diagonal elements of a matrix ${\bf{A}}$, and  ${\mathop{\rm diag}\nolimits} \left( {\bf{x}} \right)$ is a diagonal matrix with ${\bf{x}}$  being its diagonal elements.  ${x_j}$ is the  $j$th entry of a vector ${\bf{x}}$.  ${{\bf{A}}_i}$, ${{\bf{A}}^j}$  and ${A_{ij}}$ are the $j$th column, $i$th row and $(i,j)$th entry of a matrix ${\bf{A}}$, respectively.   ${x'}\left( \theta  \right)$, ${x''}\left( \theta  \right)$ are the 1st, 2nd-order derivative of $x\left( \theta  \right)$ with respect to $\theta $ , respectively. $ \odot $  denotes the Hadamard product.  ${{\bf{I}}_M}$ stands for an $M \times M$ identity matrix.

The remainder of the paper is organized as follows. We present the measurement models in Section II, and review CS-based and ML for DOA estimation in Section III. We analyze the convexity of the DML function in Section IV, and derive our CSDML algorithms in Section V. Numerical examples appear in Section VI. We give our conclusions in Section VII.

\section{MEASUREMENT MODELS}

Consider $K$ narrowband far-field signals impinging on an array composed of $M(M>K)$  sensors. The array output can be written as\cite{Krim1996}
\begin{align}\label{eqn:DOA}
{\bf{x}}\left( t \right) = {\bf{A}}\left( {\bm{\theta}}  \right){\bf{s}}\left( t \right) + {\bf{n}}\left( t \right),t = 1,\cdots,T,
\end{align}
where ${\bf{x}}\left( t \right)$ is the measurement vector at the $t$th snapshot,
${\bf{A}}\left( {\bm{\theta}}  \right) = \left[ {{\bf{a}}\left( {{\theta _1}} \right), \cdots ,{\bf{a}}\left( {{\theta _K}} \right)} \right]$
(denoted as ${\bf{A}}$ for convenience) is the array steering matrix,
${\bf{a}}\left( {{\theta _k}} \right)$
is the steering vector corresponding to the $k$th source,
${{\bm{\theta}}} = {\left[ {{\theta _1}, \cdots ,{\theta _K}} \right]^{\rm{T}}}$
is the vector containing the DOAs of all sources, ${\theta _k}$ is the DOA of the $k$th source, ${\bf{s}}\left( t \right)$ is the vector of all signal values at the $t$th snapshot, $T$ is the total number of temporal measurements, and ${\bf{n}}\left( t \right)$ is the $M \times 1$ noise vector following the zero-mean circular complex Gaussian distribution with covariance matrix $E\{ {{\bf{n}}\left( t \right){{\bf{n}}^{\rm H}}\left( t \right)} \} = {\sigma ^2}{{\bf{I}}_M}$, ${\sigma ^2}$ is noise power. With the further assumption that
${\bf{s}}\left( t \right)$ and  ${\bf{n}}\left( t \right)$  are uncorrelated, the array covariance matrix is given by
\begin{align}\label{eqtR}
{\bf{R}} = {\bf{A}}{\bf\Sigma} {{\bf{A}}^{\mathop{\rm H}\nolimits} } + {\sigma ^2}{{\bf{I}}_M},
\end{align} where ${\bf{\Sigma }}$ is the source covariance matrix.
${\bf{R}}$ can be estimated by $\widehat {\bf{R}} \approx \frac{1}{T}\sum\nolimits_{t = 1}^T {{\bf{x}}\left( t \right){{\bf{x}}^{\rm H}}\left( t \right)}$.

\section{DOA Estimation: CS-based DOA Estimation and Maximum Likelihood Estimation}
In this section we review two kinds of methods for DOA estimation, as we will take advantage of their conclusions in our context.

\subsection{CS-based DOA Estimation}
Since DOAs are sparse in the spatial domain, the source localization problem is formulated as a sparse representation problem, or spatial compressive sensing (SCS) problem \cite{Malioutov2005}. To cast the DOA estimation problem in a SCS framework, an overcomplete representation $\bf{\Theta}$ in terms of all possible source locations was introduced. Let $ {\bf{\Theta}}  = [ {{{ \Theta }_1}, \cdots ,{{ \Theta }_N}} ]$  be a sampling grid of all source locations of interest and ${\bf{\Psi}}  = [ {{\bf{a}}( {{{ \Theta }_1}} ), \cdots ,{\bf{a}}( {{{ \Theta }_N}} )} ]$ is the array manifold matrix corresponding to the DOAs ${\bf{\Theta}}$. The number of potential source locations $N$ will typically be much greater than the number of sources  $K$ or even the number of sensors $M$. Assuming that the true source directions ${\bm{\theta}}$ are contained in $ {\bf{\Theta}}$, formulation ({\ref{eqn:DOA}}) is reformulated using an overcomplete representation as the following form,
\begin{align}{\label{eqn:SCS}}
{\bf{x}}\left( t \right) = {\bf{\Psi }}\widetilde {\bf{s}}\left( t \right) + {\bf{n}}\left( t \right).
\end{align}
where, $\widetilde {\bf{s}}\left( t \right)$ is a $N \times 1$ vector containing $K$ non-zero entries, where the $n$th element ${\widetilde {\bf{s}}_n}\left( t \right)={{\bf{s}}_k}\left( t \right)$ if ${\Theta _n} = {\theta _k}$ for $k = 1,\cdots,K$, otherwise ${\widetilde {\bf{s}}_n}\left( t \right) = 0$.

In effect, model ({\ref{eqn:SCS}}) allows us to exchange the DOA estimation problem for the problem of recovering the sparse signal $\widetilde {\bf{s}}$ from the array output ${\bf{x}}$. When the number of snapshots $T>1$, this problem is a simultaneous sparse approximation problem, which has received a lot of attention recently. And several computationally feasible methods have been presented for estimating the sparse signal, such as matching pursuit (MP), convex relaxation, sparse Bayesian learning (SBL).

\subsection{Maximum Likelihood DOA Estimation}
The ML approach is a standard technique in statistical estimation theory. The ML estimation is calculated as the values of the unknown parameters that maximize the likelihood function. This can be interpreted as selecting the set of parameters that make the observed data most probable. ML techniques for the sensor array problem have been studied by a number of researchers, see for example \cite{Stoica1990b,Pesavento2001,Li2008}. When applying the ML technique to the sensor array problem, two main methods have been considered, depending on the model assumption on the signal waveforms. When the emitter signals are modeled as Gaussian random processes, a stochastic ML (SML) method is obtained.  On the other side,if the emitter signals are modeled as unknown deterministic quantities, the resulting estimator is referred to as the deterministic ML (DML) estimator. According to \cite{Ottersten1993}, we know that the DML criterion depends on ${\bm{\theta}}$ in a simpler way than does the SML criterion. So we prefer the DML estimator in this paper.

The DML DOA estimator is shown as follow \cite{Stoica1989,Ottersten1993}
\begin{align}\label{fhihat}
{\bm{\vartheta}}  = \arg \mathop {\min }\limits_{{\bm{\theta}}}  {\mathop{\rm tr}\nolimits} \left( {{{\bf{P}}^ \bot }{\bf{R}}} \right),
\end{align}
where ${{\bf{P}}^ \bot } = {\bf{I}} - {\bf{P}}$, ${\bf{P}} = {\bf{B}}{{\bf{B}}^\dag }$, in which ${\bf{B}}$ is the array steering matrix corresponding to the DOAs estimation ${\bm{\vartheta}}  = \left[ {{\vartheta _1}, \cdots ,{\vartheta _K}} \right]$, ${{\bf{B}}^\dag } = {\left( {{{\bf{B}}^{\rm H}}{\bf{B}}} \right)^{ - 1}}{{\bf{B}}^{\rm H}}$ is the pseudo-inverse of ${\bf{B}}$. However, it is important to note that (\ref{fhihat}) is a non-linear multidimensional minimization problem, and the criterion function often possesses a large number of local minima.

All methods considered herein require a multidimensional non-linear optimization for computing the signal parameter estimates. Usually, analytical solutions are not available and one has to resort to numerical search techniques. Several optimization methods have appeared in the array processing literature, including different Newton-type techniques. It is well-known that the Newton-type method gives locally a quadratic convergence. So Newton-type algorithms for the DML techniques are described in this subsection. The estimate is iteratively calculated as
\begin{align}\label{fhik}
{{\bm{\vartheta}} ^{k + 1}} = {{\bm{\vartheta}} ^k} - {{\bf{H}}^{ - 1}}{\bm{\nabla}},
\end{align}
where ${{{\bm{\vartheta}}} ^k}$ is the estimate at iteration $k$, ${\bf{H}}$ represents the Hessian matrix of the criterion function, and $\bf{\nabla} $ is the gradient. The Hessian and gradient are evaluated at ${{{\bm{\vartheta}}} ^k}$.

The DML gradient is given by \cite{Sandkuhler1987,Stoica1989}.
\begin{align}\label{eqn:grad}
\bf{\nabla}  =  - 2\Re \left( {{\mathop{\rm diag}\nolimits} \left( {{{\bf{B}}^\dag }{\bf{R}}{{\bf{P}}^ \bot }{\bf{D}}} \right)} \right).
\end{align}
The DML Hessian matrix can be expressed as (See Appendix \ref{DerivHessian})
\begin{align}\label{eqn:Hessian}
{\bf{H}} = 2\Re \left( {\bf{C}} \right),
\end{align}
where
\begin{IEEEeqnarray}{rCl}\label{eqn:expHessian}
{\bf{C}} &=&\left( {{{\bf{D}}^{\rm H}}{{\bf{P}}^ \bot }{\bf{D}}} \right) \odot {\left( {{{\bf{B}}^\dag }{\bf{R}}{{\bf{B}}^{\dag {\rm H}}}} \right)^{\rm T}}\IEEEyesnumber\IEEEyessubnumber*\label{eqn:expH1}\\
&&-\:\left( {{{\bf{D}}^{\rm H}}{{\bf{P}}^ \bot }{\bf{R}}{{\bf{P}}^ \bot }{\bf{D}}} \right) \odot {\left( {{{\bf{B}}^\dag }{{\bf{B}}^{\dag {\rm H}}}} \right)^{\rm T}}\label{eqn:expH2}\\
&&+ \left( {{{\bf{B}}^\dag }{\bf{D}}} \right) \odot {\left( {{{\bf{B}}^\dag }{\bf{R}}{{\bf{P}}^ \bot }{\bf{D}}} \right)^{\rm T}}\label{eqn:expH3}\\
&&+ {\left( {{{\bf{B}}^\dag }{\bf{D}}} \right)^{\rm T}} \odot \left( {{{\bf{B}}^\dag }{\bf{R}}{{\bf{P}}^ \bot }{\bf{D}}} \right)\label{eqn:expH4}\\
&&- {\bf{I}} \odot {\left( {{{\bf{B}}^\dag }{\bf{R}}{{\bf{P}}^ \bot }{\bf{F}}} \right)^{\rm T}}\label{eqn:expH5},
\end{IEEEeqnarray}
where ${\bf{D}} = \left[ {{{\bf{D}}_1}, \cdots ,{{\bf{D}}_K}} \right]$, ${\bf{F}} = \left[ {{{\bf{F}}_1}, \cdots ,{{\bf{F}}_K}} \right]$, ${{\bf{D}}_i} = {{\bf{a}}'}\left( {{\vartheta _i}} \right)$, ${{\bf{F}}_i} = {{\bf{a}}''}\left( {{\vartheta _i}} \right)$.

 Since this paper needs to analyze the convexity of the DML function, it is necessary to use accurate Hessian matrix expression. For the expression of Hessians matrix of both \cite{Sandkuhler1987} and \cite{Stoica1989} is asymptotic, that of our paper differs from them.

\section{The Convexity of the Likelihood Function}
\subsection{Motivation}
Although the Newton method can quickly converge to a local extremum, the quality of the convergence point depends on the shape of the criterion function in question. If $\bf{\nabla}$ possesses several minima, the iteration must be initialized ``sufficiently close'' to the global minimum in order to prevent convergence to a local extremum.

How ``sufficiently close'' to the global minimum will the iteration converge? A sufficient condition is that the likelihood function is convex in this area near the global minimum. So we will analyze the convex of the DML function in this section.
\subsection{The Convexity of the Likelihood Function near the True DOAs}\label{subsec}
The following results about gradient and Hessian matrix will show that the likelihood function is a convex function near the true DOAs.

\begin{lemma}\label{Lemma1}
    ${{\bf{D}}^{\rm H}}{{\bf{P}}^ \bot }{\bf{D}} \succ 0$.
\end{lemma}
\begin{IEEEproof}
See Appendix {\ref{ProofLemma1}}.
\end{IEEEproof}

\begin{theorem}\label{Theorem1}
    $\mathop {\lim }\limits_{{\bm{\vartheta}} \to {{\bm{\theta}}}} \bf{\nabla}  = {\bf{0}}$, $\mathop {\lim }\limits_{{\bm{\vartheta}} \to {{\bm{\theta}}}} {\bf{H}} \succ 0$.
\end{theorem}
\begin{IEEEproof}
See Appendix {\ref{ProofTheorem1}}.
\end{IEEEproof}

\begin{remark}
Theorem \ref{Theorem1} indicates that ${{\bm{\theta}}}$ is a minimum of likelihood function. Note that DML function is a continuous function in terms of  ${{\bm{\theta}}}$, thus there is a range ${\bf{\Omega }}$ around the true DOAs ${{\bm{\theta}}}$, where  $\forall {\bm{\vartheta}} \in {\bf{\Omega }}$, ${\bf{H}} \succ 0$.
\end{remark}

\subsection{The Convex Range of the Likelihood Function}
The likelihood function is a convex function in the vicinity of the global minimum according to subsection {\ref{subsec}}. As long as the initial value is in the convex vicinity of the DML solution, the Newton iteration is feasible. It is necessary to determine the convex range of the DML function. To determine the convex range of the DML function, we need to analyze the positive definiteness of ${\bf{H}}$. Unfortunately, this  analysis is rather complicated because ${\bf{H}}$ is a multidimensional nonlinear matrix function in terms of ${\bm{\vartheta}} $. We consider a simply scene where array is a uniform linear array (ULA), which is one of the most common array. Without loss of generality, the norm of the steering vector is normalized. For arbitrary ULAs, whose sensors are located at $d_1$, $\cdots$, $d_M$, we have ${\bf{a}}\left( \theta  \right) = \frac{1}{{\sqrt M }}\exp \left( { - j\frac{{2\pi }}{\lambda }\sin \left( \theta  \right){{[\;{d_1} \cdots \;{d_M}]}^{\rm{H}}}} \right)$, ${{\bf{a}}^\prime }\left( \theta  \right) =  - j\frac{{2\pi }}{\lambda }\cos \left( \theta  \right){[\;{d_1} \cdots \;{d_M}]^{\rm{H}}} \odot {\bf{a}}\left( \theta  \right)$.

For simplicity, we assume that: (1) the number of sensor elements $M$ is large enough; (2) The interval between adjacent true DOAs ${{\bm{\theta}}}$ is large enough with fixed $M$; (3) The interval between the estimated DOAs ${\bm{\vartheta}}$ and the true DOAs is much smaller comparing to the interval between adjacent DOAs, namely $\left| {{\vartheta _i} - {\theta _i}} \right| \ll \left| {{\theta _j} - {\theta _i}} \right|$, $\left| {{\vartheta _i} - {\theta _i}} \right| \ll \left| {{\vartheta _j} - {\vartheta _i}} \right|$, $i \ne j$, where ${\vartheta _i}$ is the estimation of ${\theta _i}$. According to the assumption (1), (2) and (3), we have ${\left| {{\bf{B}}_i^{\rm H}{{\bf{A}}_i}} \right|^2} \gg {\left| {{\bf{B}}_i^{\rm H}{{\bf{A}}_j}} \right|^2} \approx 0$, ${\left| {{\bf{D}}_i^{\rm H}{{\bf{A}}_i}} \right|^2} \gg {\left| {{\bf{D}}_i^{\rm H}{{\bf{A}}_j}} \right|^2} \approx 0$, ${\left| {{\bf{D}}_i^{\rm{H}}{{\bf{D}}_i}} \right|^2} \gg {\left| {{\bf{D}}_i^{\rm{H}}{{\bf{D}}_j}} \right|^2} \approx 0$. Those expressions are shown in APPENDIX G of \cite{Stoica1989}. We can reformulate them in the form of matrix as follows
 \begin{IEEEeqnarray}{RCL}
{{\bf{B}}^{\rm H}}{\bf{A}} &\approx& \left( {{{\bf{B}}^{\rm H}}{\bf{A}}} \right) \odot {\bf{I}},\label{eqn:AAS}\\
{{\bf{D}}^{\rm H}}{\bf{A}} &\approx& \left( {{{\bf{D}}^{\rm H}}{\bf{A}}} \right) \odot {\bf{I}},\label{eqn:ADS}\\
{{\bf{D}}^{\rm H}}{\bf{D}} &\approx& \left( {{{\bf{D}}^{\rm H}}{\bf{D}}} \right) \odot {\bf{I}}.\label{eqn:DDS}
\end{IEEEeqnarray}
Based on (\ref{eqn:AAS}) (\ref{eqn:ADS}) (\ref{eqn:DDS}) and after some matrix manipulations, we further have ${{\bf{B}}^{\rm H}}{\bf{B}} = {\bf{I}}$, ${{\bf{B}}^\dag }{{\bf{B}}^{\dag {\rm H}}} = {\bf{I}}$, ${{\bf{B}}^\dag }{\bf{A}} = \left( {{{\bf{B}}^\dag }{\bf{A}}} \right) \odot {\bf{I}}$, ${{\bf{D}}^{\rm H}}{{\bf{P}}^ \bot }{\bf{A}} = \;\;\left( {{{\bf{D}}^{\rm H}}{{\bf{P}}^ \bot }{\bf{A}}} \right) \odot {\bf{I}}$.

\begin{lemma}\label{Lemma2}
${{\bf{B}}^\dag }^i{\bf{A\Sigma }}{{\bf{A}}^{\rm{H}}}{{\bf{P}}^ \bot }{{\bf{D}}_i}$ is real, where  ${{{\bf{B}}}^\dag }^i$ denotes  $i$th row vector of ${{\bf{B}}^\dag }$.
\end{lemma}
\begin{IEEEproof}
See Appendix {\ref{ProofLemma2}}.
\end{IEEEproof}

Substituting (\ref{eqtR}) into (\ref{eqn:expH3}), we have
\begin{align}\label{eqn:BDI}
(\ref{eqn:expH3}) = \left( {\left( {{{\bf{B}}^\dag }{\bf{D}}} \right) \odot {\bf{I}}} \right) \odot {\left( {{{\bf{B}}^\dag }{\bf{A}}\bf{\Sigma} {{\bf{A}}^{\rm H}}{{\bf{P}}^ \bot }{\bf{D}}} \right)^{\rm T}}.
\end{align}
From (\ref{eqn:BDI}) we know (\ref{eqn:expH3}) is an approximately diagonal matrix, whose $i$th diagonal element is
\begin{align}
{[ (\ref{eqn:expH3})]_{ii}} = ({{\bf{B}}^\dag }^i{{\bf{D}}_i})({{\bf{B}}^\dag }^i{\bf{A\Sigma }}{{\bf{A}}^{\rm{H}}}{{\bf{P}}^ \bot }{{\bf{D}}_i}).
\end{align}
 Since $\alpha_i  = {{ - j2\pi \cos \left( {{\vartheta _i}} \right)} \mathord{\left/
 {\vphantom {{ - j2\pi d\cos \left( {{\vartheta _i}} \right)} \lambda }} \right.
 \kern-\nulldelimiterspace} \lambda }$ is a pure imaginary number, ${\bf{B}}_i^{\rm{H}}{{\bf{D}}_i} = \frac{{{\alpha _i}}}{M}\sum\nolimits_{k = 1}^M {{d_k}} $ is a pure imaginary number. Note that ${{\bf{B}}^\dag }^i{{\bf{D}}_i} \approx {\bf{B}}_i^{\rm H}{{\bf{D}}_i}$, ${{\bf{B}}^\dag }^i{{\bf{D}}_i}$ is an approximately imaginary number. Combining Lemma \ref{Lemma2}, ${[ (\ref{eqn:expH3})]_{ii}}$ is an approximately imaginary number. Namely, (\ref{eqn:expH3}) is an approximately diagonal matrix with approximately imaginary numbers being its diagonal elements. So,
\begin{align}\label{eqn:H3S}
\Re \left( {(\ref{eqn:expH3})} \right) \approx {\bf{0}}.
\end{align}

For similar reasons,
\begin{align}\label{eqn:H4S}
\Re \left( {(\ref{eqn:expH4})}\right) \approx {\bf{0}}.
\end{align}

Note that the formula (\ref{eqn:expH5}) includes ${{\bf{F}}_k}$, the second derivative of the steering vector ${\bf{a}}\left( {{\vartheta _k}} \right)$, it is more difficult to take this term into the analysis of formula (\ref{eqn:expHessian}). We will ignore this term without proof. There are two underlying reasons for doing like this. Firstly, the analysis of the positive definiteness of $\bf{H}$ will become easier. Secondly, the influence of neglecting (\ref{eqn:expH5}) is nearly negligible in terms of the positive definiteness of $\bf{H}$ in practice. Therefore, after using the formula (\ref{eqn:H3S})-(\ref{eqn:H4S}) and neglecting (\ref{eqn:expH5}), ${\bf{H}} = 2\Re \left( {\bf{C}} \right)$ can be approximated as follows
\begin{IEEEeqnarray}{rCl}
{\bf{C}} & \approx & ( \ref{eqn:expH1})+(\ref{eqn:expH2})
\nonumber\\
& = &\left( {{{\bf{D}}^{\rm H}}{{\bf{P}}^ \bot }{\bf{D}}} \right) \odot {\left( {{{\bf{B}}^\dag }{\bf{A\Sigma }}{{\bf{A}}^{\rm H}}{{\bf{B}}^{\dag {\rm H}}}} \right)^{\rm T}} \nonumber\\
&\:&-\left( {{{\bf{D}}^{\rm H}}{{\bf{P}}^ \bot }{\bf{A\Sigma }}{{\bf{A}}^{\rm H}}{{\bf{P}}^ \bot }{\bf{D}}} \right) \odot {\left( {{{\bf{B}}^\dag }{{\bf{B}}^{\dag {\rm H}}}} \right)^{\rm T}} \nonumber\\
& = &\left( {\left( {{{\bf{D}}^{\rm H}}{{\bf{P}}^ \bot }{\bf{D}}} \right) \odot {\bf{I}}} \right) \odot {\left( {\left( {\left( {{{\bf{B}}^\dag }{\bf{A}}} \right) \odot {\bf{I}}} \right){\bf{\Sigma}} {{\left( {\left( {{{\bf{B}}^\dag }{\bf{A}}} \right) \odot {\bf{I}}} \right)}^{\rm H}}} \right)^{\rm T}} \nonumber\\
&\:&-\left( {\left( {\left( {{{\bf{D}}^{\rm H}}{{\bf{P}}^ \bot }{\bf{A}}} \right) \odot {\bf{I}}} \right){\bf{\Sigma }}{{\left( {\left( {{{\bf{D}}^{\rm H}}{{\bf{P}}^ \bot }{\bf{A}}} \right) \odot {\bf{I}}} \right)}^{\rm H}}} \right) \odot {\bf{I}}.
\end{IEEEeqnarray}
So ${\bf{C}}$ is an approximately diagonal matrix, whose  $i$th diagonal element is
\begin{IEEEeqnarray}{rCl}\label{eqn:Cii}
{{\bf{C}}_{ii}}&=&{\Sigma _{ii}}\left( {{{\bf{B}}^\dag }^i{{\bf{A}}_i}} \right){\left( {{{\bf{B}}^\dag }^i{{\bf{A}}_i}} \right)^{\rm H}}{\bf{D}}_i^{\rm H}{{\bf{P}}^ \bot }{{\bf{D}}_i}\nonumber\\
&\:&-{\Sigma _{ii}}{\bf{D}}_i^{\rm H}\left( {{{\bf{P}}^ \bot }{{\bf{A}}_i}} \right){\left( {{{\bf{P}}^ \bot }{{\bf{A}}_i}} \right)^{\rm H}}{{\bf{D}}_i}\nonumber\\
& = &{\Sigma _{ii}}{\beta _i}{\bf{D}}_i^{\rm H}{{\bf{P}}^ \bot }{{\bf{D}}_i} - {\Sigma _{ii}}\left\| {{{\bf{P}}^ \bot }{{\bf{A}}_i}} \right\|_F^2{\bf{D}}_i^{\rm H}\left( {{{\bf{e}}_i}{\bf{e}}_i^{\rm H}} \right){{\bf{D}}_i}\nonumber\\
& = &{\Sigma _{ii}}\left( {{\beta _i}{\bf{D}}_i^{\rm H}{{\bf{P}}^ \bot }{{\bf{D}}_i} - \left( {1 - {\beta _i}} \right){\bf{D}}_i^{\rm H}\left( {{{\bf{e}}_i}{\bf{e}}_i^{\rm H}} \right){{\bf{D}}_i}} \right),
\end{IEEEeqnarray}
where ${\beta _i} = {\left| {{\bf{B}}_i^{\rm H}{{\bf{A}}_i}} \right|^2}$, ${{\bf{e}}_i} \buildrel \Delta \over = \frac{{{{\bf{P}}^ \bot }{{\bf{A}}_i}}}{{{{\left\| {{{\bf{P}}^ \bot }{{\bf{A}}_i}} \right\|}_2}}}$, $\left\| {{{\bf{P}}^ \bot }{{\bf{A}}_i}} \right\|_2^2 =1 - {\left| {{\bf{B}}_i^{\rm{H}}{{\bf{A}}_i}} \right|^2}= 1 - {\beta _i}$.
\begin{lemma}\label{Lemma3}
${{\bf{P}}^ \bot } - {{\bf{e}}_i}{\bf{e}}_i^{\rm H} \succeq 0$.
\end{lemma}
\begin{IEEEproof}
See Appendix {\ref{ProofLemma3}}.
\end{IEEEproof}

According to Lemma \ref{Lemma3}, we have
\begin{IEEEeqnarray}{rCl}\label{eqn:DPD}
{\bf{D}}_i^{\rm H}{{\bf{P}}^ \bot }{{\bf{D}}_i} - {\bf{D}}_i^{\rm H}\left( {{{\bf{e}}_i}{\bf{e}}_i^{\rm H}} \right){{\bf{D}}_i} \ge 0.
\end{IEEEeqnarray}

If ${\beta _i} \ge 0.5$, ${\beta _i} \ge 1 - {\beta _i}$. Substituting ${\beta _i} \ge 1 - {\beta _i}$, ${{\bf{\Sigma }}_{ii}} > 0$ and (\ref{eqn:DPD}) into (\ref{eqn:Cii}), we have
 ${{\bf{C}}_{ii}} \ge 0$. Note that both ${\bf{C}}$ and ${\bf{H}}$ are approximately diagonal matrices, we have the following result
\begin{align} \label{eqn:conclu}
{\bf{H}} \succeq 0,when\;{\left| {{\bf{B}}_i^{\rm H}{{\bf{A}}_i}} \right|^2} \ge 0.5,i = 1,2, \cdots ,K.
\end{align}

\begin{remark}

Since we use the approximation (\ref{eqn:AAS})(\ref{eqn:ADS})(\ref{eqn:DDS}) and neglect  (\ref{eqn:expH5}) to obtain (\ref{eqn:conclu}),  the convex range given by (\ref{eqn:conclu}) is a rough range. Even so, the result is often useful to roughly determine convex range of the DML function. In fact, the rough range  accords well with the real range, as long as the hypothesis conditions are fully satisfied. This will be shown by simulations. The significance of the result lies in its giving some guidance for the iterative algorithm initialization in practice. As long as the initial value for the iterative algorithm is in the above convex range, the algorithm will converge to the global minimum of the DML function. The approximate convex range is acquired through solving the inequation ${\left| {{\bf{B}}_i^{\rm H}{{\bf{A}}_i}} \right|^2} \ge 0.5$ using the numerical method. In this way, the range is slightly different when ${\theta _i}$ varies. The range is almost same as long as ${\theta _i}$ is not a big angle. So, we take the width where ${\theta _i}=0$  as the range. Namely, the approximately convex range is about half of 3dB (half-power) beamwidth $\rm{BW}_{0.5}$ of the array around the true DOAs.
\end{remark}

\section{CSDML: Off-Grid DOA Estimation Based on the Convexity of DML Function}
CS-based DOA estimation methods based on on-grid model can provide estimated DOAs which are on its nearest grid point since the estimated DOAs are constrained on the grid. And the estimated DOAs can be uesd to initialize Newton iteration. Since the result (\ref{eqn:conclu}) indicates that by and large the convex range of the likelihood function is half of the 3dB beamwidth of the array, we can set grid interval $r$ for the CS-based DOA estimation as
\begin{align} \label{eqn:r}
r = \gamma \frac{{ \rm{BW}_{0.5} }}{2},
\end{align}
where $\gamma  < 1$ is a small regulation parameter. When the DOAs are more adjacent, $\gamma $ is set to a smaller number. We recommend to set $\gamma  = 0.5$ usually since it can give an accurate yet fast DOA estimation. Then the DOAs ${{\bm{\theta}}}_c$ estimated by the CS-based DOA estimation methods are most likely in the convex range of the DML function so that Newton iteration always converge to the global minimum rather than a local extremum.

Based on the above ideas, we propose a class of algorithms for off-grid DOA estimation, outlined in Table {\ref{Alg1}}. The proposed algorithms consist of two stages. By setting the grid interval according to our result (\ref{eqn:conclu}), CS-based DOA estimation approaches obtain coarse DOA estimates in the first stage. Next, we initialize the Newton type iteration with these estimated DOAs to realize the DML estimation.

\begin{table}[!t]
    \renewcommand{\arraystretch}{1.3}
    \caption{\textbf{Algorithm CSDML}}\label{Alg1}
    \label{table_example}
    \centering
    \begin{tabularx}{8.5cm}{lX}
        \toprule
         1)&Set grid interval $r$ according to (\ref{eqn:conclu}) and (\ref{eqn:r}), obtain ${\bf{\Theta}} $ and $\bf{\Psi} $;\\
         2)&Calculate ${{\bm{\theta}}_c}$ through the on-grid methods such as MP, convex relaxation, SBL and so on;\\
         3)&Set ${{{\bm{\vartheta}}}^0} = {{{\bm{\theta}}}_c}$ for Newton iteration;\\
         4)&At ${{{\bm{\vartheta}}} ^k}$, calculate $\bf{\nabla}$  and ${\bf{H}}$ according to (\ref{eqn:grad}) and (\ref{eqn:Hessian}), respectively;\\
         5)&Update ${{{\bm{\vartheta}}} ^{k + 1}}$ according to (\ref{fhik});\\
         6)&Check the convergence criterion for the $k$th iteration. Terminate the process when ${\left\| {{{{\bm{\vartheta}}} ^{k + 1}} - {{{\bm{\vartheta}}} ^k}} \right\|_2} \le \tau $ or the maximum number of iterations is reached, where $\tau $ is a predefined small value. Otherwise, return to step (4) and continue the whole process again.\\
        \bottomrule
    \end{tabularx}
\end{table}

\begin{remark}
It is worthy to note that all kinds of DOA estimation methods can be applied in the first stage of the proposed approach, including the CS-based methods, such as MP, convex relaxation, SBL. As long as their DOA estimates are already in the convex vicinity of the DML solution, the Newton type iteration will converge to the global minimum of the DML function and achieve the same performance in the final. We can choose the computationally effective methods in practice. Usually, OMP is decidedly less costly than all of these methods.
\end{remark}

We assume $N \gg M \gg K$, where $N$ denotes the number of sampling grid. Note that each iteration of both M-BP and M-SBL can also be computed in ${\mathcal{O}(NM^2)}$ using the implementation as \cite{Wipf2007}. After SVD is introduced, each M-OMP iteration can be computed in ${\mathcal{O}(NMK)}$ using the implementation as \cite{Tropp2006a} and standard techniques for least-squares problems (see \cite{Golub1996}, Chapter 5, for extensive details). Each Newton iteration can be computed in ${\mathcal{O}(M^2K)}$, which is not significant compared to M-BP, M-SBL or M-OMP iteration.

\section{Simulation}
This section includes two parts. Firstly, we compare the convex rough range of ours with convex exact range of the DML function. Secondly, we present the numerical simulation results to illustrate the performance of the proposed algorithm. In our experiments, we consider $K = 2$ narrowband far-field non-coherent signals with the equal power impinging on a ULA composed of $M$ sensors which are separated by a half wavelength of the signal. The SNR is defined as  ${\rm{SN}}{{\rm{R}}_i} = 10{\log _{10}}(E({\left| {{{\bf{s}}_i}(t)} \right|^2})/{\sigma ^2})$, where ${\rm{SN}}{{\rm{R}}_i}$  denotes $i$th signal's SNR. The RMSE is defined as ${\rm{RMSE}} = \sqrt {\frac{1}{{{N_m}K}}\sum\nolimits_{i = 1}^{{N_m}} {\sum\nolimits_{k = 1}^K {{{(\theta _k^i - \widehat \theta _k^i)}^2}} } } $, where the superscript $i$ refers to the $i$th trial, ${N_m}$ denotes the number of Monte Carlo tests. The simulations are performed using MATLAB2012B running on an Intel Core 2 Duo, 3.20 GHz processor with 3 GB of memory, under Windows7 32bit.
\subsection{Convexity of the Likelihood Function}\label{subsec2}
In order to measure the similarity of the range based on the (\ref{eqn:conclu}) and the exact one, We define two measures as
\begin{align}
\rm{IRR} = \sqrt[K]{{{{M\left( {{\Omega _I}} \right)} \mathord{\left/
 {\vphantom {{M\left( {{\Omega _I}} \right)} {M\left( {{\Omega _R}} \right)}}} \right.
 \kern-\nulldelimiterspace} {M\left( {{\Omega _R}} \right)}}}},
\end{align}
\begin{align}
\rm{IAR} = \sqrt[K]{{{{M\left( {{\Omega _I}} \right)} \mathord{\left/
 {\vphantom {{M\left( {{\Omega _I}} \right)} {M\left( {{\Omega _A}} \right)}}} \right.
 \kern-\nulldelimiterspace} {M\left( {{\Omega _A}} \right)}}}},
\end{align}
where, ${\Omega _R}$ represents a set, which subjects to $\forall {\bm{\vartheta}} \in {\Omega _R}$, ${\bf{H}} \succ 0$ and is obtained from experiment.  ${\Omega _A}$ represents another set, which approximately subjects to $\forall {\bm{\vartheta}}  \in {\Omega _A}$, ${\bf{H}} \succ 0$, and is obtained from (\ref{eqn:conclu}). And  ${\Omega _I} = {\Omega _R} \cap {\Omega _A}$. $M\left( \Omega  \right)$ represents the measure of $\Omega $. Taking $K$-th root of the ratio is to eliminate the influence of the dimension $K$. Obviously,  $0 \le \rm{IRR} \le 1$, $0 \le \rm{IAR} \le 1$. When ${\Omega _A}$ is closer to ${\Omega _R}$, both IRR and IAR  are closer to 1, and it means that the estimated range and the actual range is consistent. Otherwise, they are closer to 0.

In the simulations, to acquire the ${\Omega _R}$, firstly, we divide the area near the real DOAs into uniformly-spaced grid, then we calculate the eigenvalues of ${\bf{H}}$ corresponding to the value of the DOAs grid. If the eigenvalues of ${\bf{H}}$ at a grid are all non-negative, the ML function is convex at this grid, which belongs to ${\Omega _R}$. We regard the number of elements of ${\Omega _R}$ as $M\left( {{\Omega _R}} \right)$. We can easily obtain ${\Omega _A}$ and $M\left( {{\Omega _A}} \right)$ through the 3dB beamwidth $\rm{BW}_{0.5}$.

\begin{figure}[t]
\centering
\includegraphics[width=88mm]{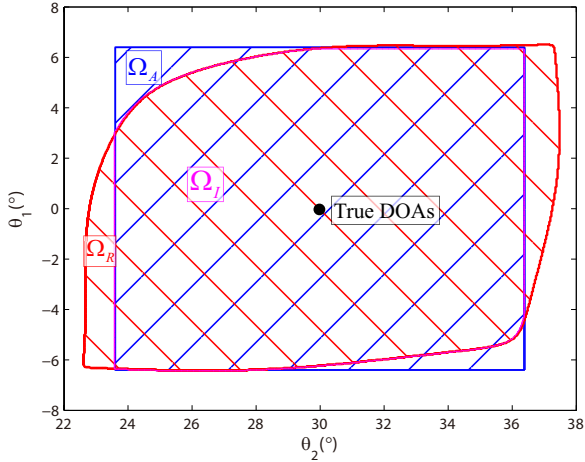}
\caption{
The rough and exact convex range of the DML function.
}
\label{figS1-1}
\end{figure}

We carry out 3 experiments in this subsection. The number of snapshots is fixed at $T = 200$ in each experiment. 500 Monte Carlo trials are simulated except the first one. Fig.{\ref {figS1-1}} is the result of an experiment with SNR = 10 dB, $M=8$ and ${{\bm{\theta}}}  = {[ { {{0}^\circ }\;{{30}^\circ }} ]^{\rm T}}$. $\rm{BW}_{0.5}\approx 12.8^\circ$ when $M=8$. It intuitively shows the convex rough range and the exact range of the DML function. Despite that the rough range based on (\ref{eqn:conclu}) is slightly different, two ranges still agree well.

\begin{figure}[t]
\centering
\includegraphics[width=88mm]{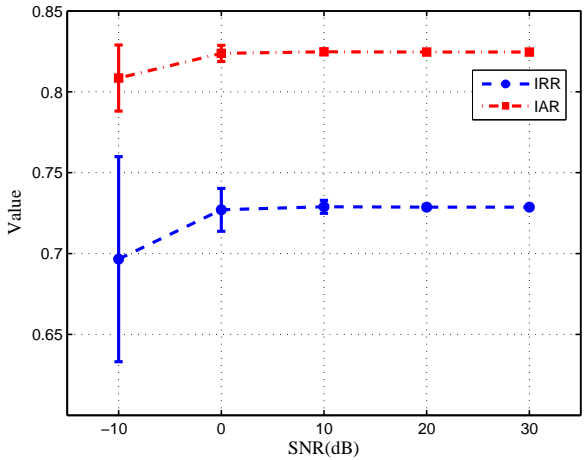}
\caption{
IRR and IAR versus SNR.
}
\label{figS1-2}
\end{figure}

Fig.{\ref {figS1-2}} depicts the IRR and IAR versus SNR, with ${{\bm{\theta}} } = {\left[ { - {{7.5}^\circ }\;{{7.5}^\circ }} \right]^{\rm T}}$ and $M=8$. It is shown that the mean of IRR and IAR slightly increase with the increasing of SNR, but nearly remains 0.73 and 0.84, respectively. As long as (\ref{eqtR}) is satisfied strictly, the convex rough is always invariable. But IRR and IAR fluctuate more remarkably around the mean value with low SNR.

\begin{figure}[t]
\centering
\includegraphics[width=88mm]{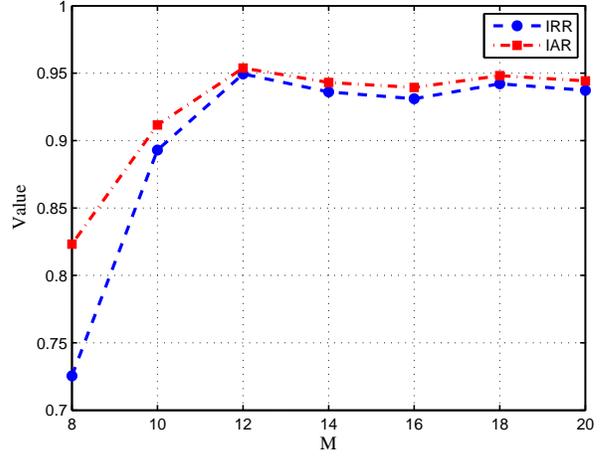}
\caption{
ICR and IAR versus the number of sensors.
}
\label{figS1-4}
\end{figure}

Fig.{\ref {figS1-4}} shows the IRR and IAR versus the number of sensors $M$ , with SNR fixed at 10 dB, DOA fixed at $[-10^\circ, 10^\circ]$. Fig.{\ref {figS1-4}} shows that by and large the mean of IRR and IAR increases with the number of sensors. Namely, the convex rough range and the exact range of the DML function agree better with more sensors. In Fig.{\ref {figS1-4}}, the means of both IRR and IAR  are about 0.93 when $M\geq12$. Namely, the convex rough range and the exact range of the DML function agree very well when DOA  interval is larger than $2.5\rm{BW}_{0.5}$ ($\rm{BW}_{0.5}\approx 8^\circ$ when $M=12$).

\subsection{DOA Estimation Performance}
In this subsection, we consider the uniform sampling grid as $\left\{ { - {{90}^\circ }, - {{90}^\circ } + r, - {{90}^\circ } + 2r, \cdots ,{{90}^\circ }} \right\}$, with $r$ being the grid interval. We fix the number of sensors of ULA at $M = 8$. We set the parameters of OGSBI-SVD  as \cite{Yang2013}. It's worth pointing out that it is necessary to set the origin at the middle point of the ULA to reduce the approximation error for OGSBI-SVD, however is unnecessary for our proposed algorithms. We assume that $K$ is known in our simulation. Note that all kinds of the CS-based methods are optional in the first stage of our proposed approach. In our simulations, we take OMP \cite{Tropp2006a} and SBL \cite{Wipf2007} as representatives of on-grid model based methods. There are two underlying reasons for doing like this. Firstly, OMP is computationally effective method. Secondly, both SBL and OGSBI-SVD are from a Bayesian perspective. We do not choose $\ell_{1}$-SVD because it costs too much usually. Since OGSBI-SVD uses the SVD to reduce the computational workload of the signal recovery process and the sensitivity to noise. For the sake of fair, the SVD is introduced for OMP and SBL, too. We compare OGSBI-SVD, SBL, OMP with our proposed algorithms ,CSDML(SBL) and CSDML(OMP), in terms of RMSE and computational time with respect to SNR, snapshots and grid interval, where CSDML(SBL) and CSDML(OMP) denote our proposed algorithms using SBL and OMP in the first stage, respectively. 1000 Monte Carlo trials for each experiment are simulated in this subsection.

It should be noted that there exists a Lower Bound for the RMSE of the on-Grid methods (GLB) regardless of the SNR since the best DOA estimate that those methods can obtain is the grid point nearest to the true DOA. In fact, the lower bound is shared among all on-grid model based methods including $\ell_{1}$-SVD, SBL, OMP and so on. In the case of the uniformly distributed DOA, the lower bound is ${\rm{GLB}} = \frac{r}{{2\sqrt 3 }}$. In the case of the fixed DOA, the lower bound is ${\rm{GLB}} = \sqrt {\frac{1}{K}\sum\nolimits_{k = 1}^K {{{({\theta _k} - {{\widehat \theta }_k})}^2}} } $, where $\widehat {{\theta}}_k$ is the grid point nearest to ${\theta _k}$.

\begin{figure}[t]
\centering
\includegraphics[width=88mm]{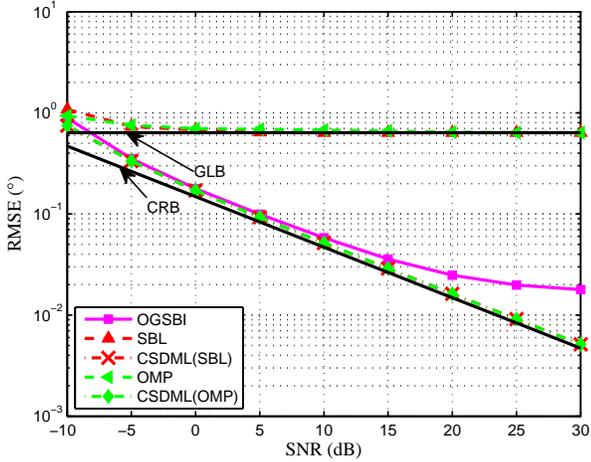}
\caption{RMSE of DOA estimates versus SNR.}
\label{figS2-1}
\end{figure}

\begin{figure}[t]
\centering
\includegraphics[width=88mm]{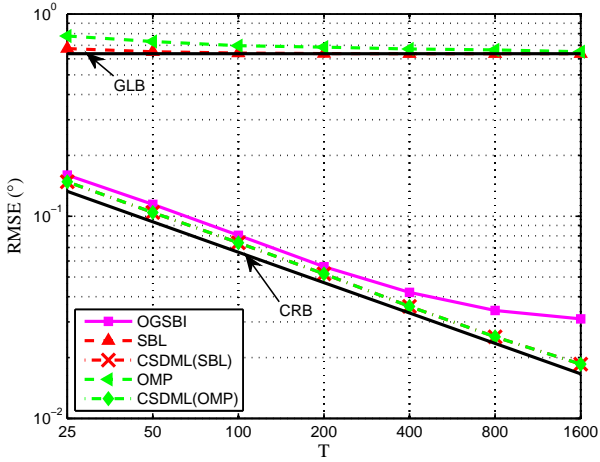}
\caption{RMSE of DOA estimates versus the number of snapshots.}
\label{figS2-2}
\end{figure}

In the first two experiments, we consider that ${{\bm{\theta}}}  = \left[ {{{2.37}^\circ },{{30.82}^\circ }} \right]$ and $r = {2^\circ }$, which is equal to that  $\lambda$ is set 0.313 since $\rm{BW}_{0.5} \approx 12.8^\circ$ for $M = 8$. Fig.{\ref {figS2-1}} depicts the RMSE versus SNR, with the number of snapshots fixed at $T= 200$, and Fig.{\ref {figS2-2}} shows the RMSE versus the number of snapshots, with SNR fixed at 10 dB. It is shown that the performance of the off-grid based methods, OGSBI-SVD and CSDML, are superior to that of the on-grid method, SBL and OMP, in terms of RMSE in Fig.{\ref {figS2-1}} and {\ref {figS2-2}}. The performance of the on-grid model will degrade due to discretization error when the true DOAs are beyond the fixed sampling grid. Fig.{\ref {figS2-1}} and {\ref {figS2-2}} show that although SBL and OMP have different performance, CSDML(SBL) and CSDML(OMP) have same performance, which outperforms OGSBI-SVD under the same simulation conditions, especially with high SNR or large snapshots, where OGSBI-SVD has the constant performances, however both CSDML(SBL) and CSDML(OMP) are almost able to achieve the CRB.

\begin{figure}[t]
\centering
\includegraphics[width=88mm]{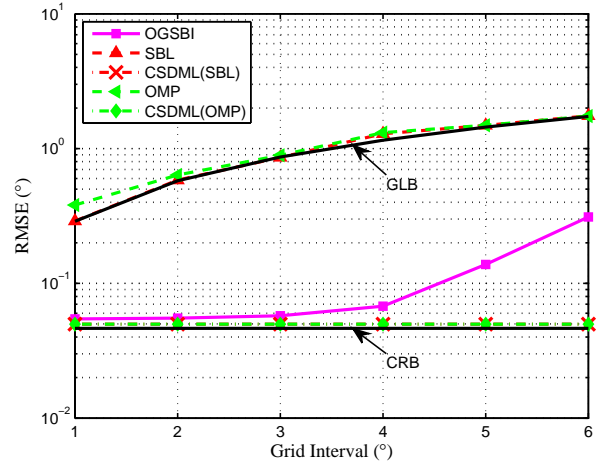}
\caption{RMSE of DOA estimates versus the grid interval.% (SNR = 10 dB, $T=200$).
}
\label{figS2-3}
\end{figure}

The sensitivity to grid interval of different methods is studied in the third experiment. The SNR and the number of snapshots are fixed at 10 dB and 200, respectively. When $M=8$, we have $\rm{BW}_{0.5} \approx 12.8^{\circ}$. Therefore the grid interval $r$ is selected from $1^\circ$ to $6^\circ$ with an interval of $1^{\circ}$. It is equal to that $\lambda$ is set from 0.156 to 0.938 with an interval of 0.156. In each trial, $K=2$ directions of $\theta_1$, $\theta_2$ are uniformly generated within direction intervals $[-3^\circ, 3^\circ]$ and $[27^\circ, 33^\circ]$ respectively. The RMSE versus different grid intervals is depicted in Fig.{\ref {figS2-3}}, which shows that CSDML has superior performance than OGSBI-SVD. This simulation also demonstrates that CSDML is more robust than OGSBI-SVD, even in the case of very coarse grid. Table {\ref {table_time}} presents the averaged CPU times of OGSBI-SVD and CSDML (excluding the SVD process) with respect to $r$. Their CPU times decrease as the grid gets coarser for both OGSBI-SVD and CSDML. However, CSDML, especially CSDML(OMP), is many times faster than OGSBI-SVD especially in the case of very fine grid interval. Additionally, the CPU times of Newton iteration are much less comparing with SBL or OGSBI-SVD, for Newton iteration gives locally a quadratic convergence. The table also shows that CSDML(OMP) is less costly than CSDML(SBL) since OMP is less costly than SBL.

\begin{table}[!t]
\renewcommand{\arraystretch}{1.3}
\caption{Averaged CPU times of OGSBI-SVD and CSDML with respect to $r$. Time unit: \textup{sec}.}
\label{table_time}
\centering
\begin{tabular}{ccccccc}
\toprule
$r$&$1^\circ$&$2^\circ$&$3^\circ$&$4^\circ$&$5^\circ$&$6^\circ$\\\midrule
OGSBI-SVD&1.1e+0&1.3e-1&5.1e-2&3.3e-2&2.8e-2&2.1e-2\\\hline
SBL&1.6e-1&1.6e-2&7.7e-3&6.1e-3&5.0e-3&3.8e-3\\
DML&5.0e-4&6.0e-4&5.9e-4&6.2e-4&6.2e-4&6.3e-4\\
CSDML(SBL)&1.6e-1&1.7e-2&8.3e-3&6.7e-3&5.6e-3&4.4e-3\\\hline
OMP&4.3e-4&3.4e-4&3.4e-4&3.2e-4&3.2e-4&3.2e-4\\
DML&5.5e-4&5.6e-4&5.9e-4&6.2e-4&6.6e-4&6.3e-4\\
CSDML(OMP)&9.8e-4&9.0e-4&9.3e-4&9.4e-4&9.8e-4&9.5e-4\\
\bottomrule
\end{tabular}
\end{table}

\section{Conclusions}

In this work, we analyzed the convexity of the DML function in the vicinity of the global solution and found out a rough range with large array, which is the half of 3dB beamwidth around the true DOAs to generally guarantee the DML function convex. It to some extent answered the question that how close it should be to the global minimum to make sure the iteration feasible. After that, we proposed algorithms for off-grid DOA estimation based on the convexity of the DML function. We illustrated by simulations that the approximately convex range agrees well with the exact convex range of the DML function with large array. The simulations demonstrated that the proposed approaches outperform standard CS methods based on on-grid model and the OGSBI-SVD based on off-grid model. The performance of the proposed approaches is close to the CRB, and exceeds that of the OGSBI-SVD, especially with high SNR and large snapshots. Moreover, the proposed approaches are robust in the case of large grid interval, whereas the OGSBI-SVD is not. Besides, the proposed approaches are many times faster than OGSBI-SVD in all case of grid interval.

From a new perspective, this work provided a computationally efficient algorithm framework including initial value calculation and Newton iteration to realize DML estimation. Unlike other approaches, our algorithms do not need a global/multidimensional search and the convex analysis of the DML function would guarantee the Newton iteration convergence. This algorithm frame can be applied to DOA estimation, frequency estimation and so on.

\appendices
\section{Derivation of the Hessian Matrix}\label{DerivHessian}

For convenience, the following matrices are defined as
\begin{IEEEeqnarray}{RCL}
{{\bf{E}}_i} \buildrel \Delta\over=&& \frac{{\partial {\bf{B}}}}{{\partial {\vartheta _i}}} = \left[ {0\;{{\bf{D}}_i}\;0} \right],\IEEEnonumber\\
{{\bf{G}}_{i,j}} \buildrel \Delta\over =&& \frac{{{\partial ^2}{\bf{B}}}}{{\partial {\vartheta _i}\partial {\vartheta _j}}} = \left\{ {\begin{array}{*{20}{c}}
{\left[ {{\bf{0}}\;{{\bf{F}}_i}\;{\bf{0}}} \right],j = i}\\
{\;\;\;\;{\bf{0}}\;\;\;\;\;,j \ne i}
\end{array}} \right..\IEEEnonumber
\end{IEEEeqnarray}
According to the results in \cite{Golub1973,Viberg1991}, we have
\begin{IEEEeqnarray}{rCl}
\frac{{{\partial ^2}{\bf{P}}}}{{\partial {\vartheta _i}\partial {\vartheta _j}}} &=& 2{\mathop{\Re}\nolimits} \left( { - {{\bf{P}}^ \bot }{{\bf{E}}_j}{{\bf{B}}^\dag }{{\bf{E}}_i}{{\bf{B}}^\dag }} \right.\IEEEnonumber\\
&&~~~~~ + {{\bf{P}}^ \bot }{{\bf{E}}_i}{{\bf{B}}^\dag }{{\bf{B}}^{\dag {\rm H}}}{\bf{E}}_j^{\rm H}{{\bf{P}}^ \bot } \IEEEnonumber\\
&&~~~~~ - {{\bf{B}}^{\dag {\rm H}}}{\bf{E}}_j^{\rm H}{{\bf{P}}^ \bot }{{\bf{E}}_i}{{\bf{B}}^{\rm H}} \IEEEnonumber\\
&&~~~~~ - {{\bf{P}}^ \bot }{{\bf{E}}_i}{{\bf{B}}^\dag }{{\bf{E}}_j}{{\bf{B}}^\dag } \IEEEnonumber\\
&&~~~~~\left. { + {{\bf{P}}^ \bot }{{\bf{G}}_{i,j}}{{\bf{B}}^\dag }} \right). \IEEEnonumber
\end{IEEEeqnarray}
Therefore, we have
\begin{IEEEeqnarray}{rCl}\label{eqn:Hij}
 {{{H}}_{ji}} &=& \frac{{{\partial ^2}{\mathop{\rm tr}\nolimits} \left( {{{\bf{P}}^ \bot }{\bf{R}}} \right)}}{{\partial {\vartheta _i}\partial {\vartheta _j}}} =  - \frac{{{\partial ^2}{\mathop{\rm tr}\nolimits} \left( {{\bf{PR}}} \right)}}{{\partial {\vartheta _i}\partial {\vartheta _j}}}\IEEEnonumber\\
&=& 2{\mathop{\Re}\nolimits} \left[ {{\mathop{\rm tr}\nolimits} \left( { + {{\bf{B}}^\dag }{\bf{R}}{{\bf{B}}^{\dag {\rm H}}}{\bf{E}}_j^{\rm H}{{\bf{P}}^ \bot }{{\bf{E}}_i}} \right.} \right. \IEEEnonumber\\
&&~~~~~~~~~ -{{\bf{B}}^\dag }{{\bf{B}}^{\dag {\rm H}}}{\bf{E}}_j^{\rm H}{{\bf{P}}^ \bot }{\bf{R}}{{\bf{P}}^ \bot }{{\bf{E}}_i}\IEEEnonumber\\
&&~~~~~~~~~ + {{\bf{B}}^\dag }{\bf{R}}{{\bf{P}}^ \bot }{{\bf{E}}_j}{{\bf{B}}^\dag }{{\bf{E}}_i}\IEEEnonumber\\
&&~~~~~~~~~ + {{\bf{B}}^\dag }{\bf{R}}{{\bf{P}}^ \bot }{{\bf{E}}_i}{{\bf{B}}^\dag }{{\bf{E}}_j}\IEEEnonumber\\
&&~~~~~~~~~ \left. {\left. { - {{\bf{B}}^\dag }{\bf{R}}{{\bf{P}}^ \bot }{{\bf{G}}_{i,j}}} \right)} \right].
\end{IEEEeqnarray}
By comparison, (\ref{eqn:Hessian}) is the matrix form of (\ref{eqn:Hij}).

\section{Proof of Lemma \ref{Lemma1}}\label{ProofLemma1}
Suppose arbitrary vector ${\bf{x}} \ne {\bf{0}}$.
${{\bf{P}}^ \bot } = {\bf{I}} - {\bf{B}}{{\bf{B}}^\dag }$ is the orthogonal projector matrix and the rank of ${\bf{B}}$ is $K$ , therefore its eigenvalues are composed of ones and zeros, and the number of them is $M - K$ and $K$ , respectively. Thus ${\lambda _{\min }}\left( {{{\bf{P}}^ \bot }} \right) = 0$.

Obviously, ${{\bf{D}}^{\rm H}}{{\bf{P}}^ \bot }{\bf{D}}$ is a Hermitian matrix. And according to the definition of the Rayleigh quotient and Min-max theorem in \cite{Horn1985}, we have
\begin{align}
{{\bf{x}}^{\rm{H}}}{{\bf{D}}^{\rm{H}}}{{\bf{P}}^ \bot }{\bf{D}}{\bf{x}} = {\left( {{\bf{D}}{\bf{x}}} \right)^{\rm{H}}}{{\bf{P}}^ \bot }\left( {{\bf{D}}{\bf{x}}} \right) \ge {\lambda _{\min }}\left( {{{\bf{P}}^ \bot }} \right){\left\| {{\bf{D}}{\bf{x}}} \right\|^2}.\nonumber
\end{align}
If and only if ${\bf{D}}{\bf{x}}$ is the eigenvector of ${{\bf{P}}^ \bot }$  corresponding to the smallest eigenvalue 0, ${{\bf{x}}^{\rm H}}{{\bf{D}}^{\rm H}}{{\bf{P}}^ \bot }{\bf{D}}{\bf{x}} = 0$. Otherwise, ${{\bf{x}}^{\rm H}}{{\bf{D}}^{\rm H}}{{\bf{P}}^ \bot }{\bf{D}}{\bf{x}} > 0$ .

Since ${{\bf{P}}^ \bot }{\bf{B}} = {\bf{0}}$, then ${\bf{a}}\left( {{\vartheta _i}} \right)$ , column vector of ${\bf{B}}$, is eigenvector of ${{\bf{P}}^ \bot }$ corresponding to eigenvalue 0. If and only if ${\bf{D}}{\bf{x}} \in R({\bf{B}})$, where $R({\bf{B}})$ denotes the column space of $\bf{B}$, ${\left( {{\bf{D}}{\bf{x}}} \right)^{\rm H}}{{\bf{P}}^ \bot }\left( {{\bf{D}}{\bf{x}}} \right) = 0$. However, ${{\bf{D}}_i} = {{\bf{a}}'}\left( {{\vartheta _i}} \right)$ and ${{\bf{B}}_i} = {\bf{a}}\left( {{\vartheta _i}} \right)$, thus ${\bf{D}}{\bf{x}} \in R({\bf{B}})$ is usually not satisfied, and ${\left( {{\bf{D}}{\bf{x}}} \right)^{\rm H}}{{\bf{P}}^ \bot }\left( {{\bf{D}}{\bf{x}}} \right) > 0$.
Namely, $\forall {\bf{x}} \ne 0$, ${\left( {{\bf{D}}{\bf{x}}} \right)^{\rm H}}{{\bf{P}}^ \bot }\left( {{\bf{D}}{\bf{x}}} \right) > 0$. Therefore ${{\bf{D}}^{\rm H}}{{\bf{P}}^ \bot }{\bf{D}} \succ 0$ .

\section{Proof of Theorem \ref{Theorem1}}\label{ProofTheorem1}
Considering (\ref{eqtR}) and after some matrix manipulations, we have
\begin{IEEEeqnarray}{RCL}
{{\bf{B}}^\dag }{\bf{R}}{{\bf{P}}^ \bot } &=& {{\bf{B}}^\dag }{\bf{A\Sigma }}{{\bf{A}}^{\rm{H}}}{{\bf{P}}^ \bot },\label{eqn:grd}\\
(\ref{eqn:expH1})+(\ref{eqn:expH2}) &=& \left( {{{\bf{D}}^{\rm H}}{{\bf{P}}^ \bot }{\bf{D}}} \right) \odot {\left( {{{\bf{B}}^\dag }{\bf{A}}{\bf{\Sigma}} {{\bf{A}}^{\rm H}}{{\bf{B}}^{\dag {\rm H}}}} \right)^{\rm T}}\nonumber\\
&&\, -\left( {{{\bf{D}}^{\rm H}}{{\bf{P}}^ \bot }{\bf{A}}{\bf{\Sigma}} {{\bf{A}}^{\rm H}}{{\bf{P}}^ \bot }{\bf{D}}} \right) \odot {\left( {{{\bf{B}}^\dag }{{\bf{B}}^{\dag {\rm H}}}} \right)^{\rm T}}.\label{eqtTH2}
\end{IEEEeqnarray}

Taking the limit of (\ref{eqn:grd}) and (\ref{eqtTH2}), respectively, and note that $\mathop {\lim }\limits_{{\bm{\vartheta}}  \to {{\bm{\theta}}}} {{\bf{A}}^{\rm{H}}}{{\bf{P}}^ \bot } = {\bf{0}}$, we have
\begin{IEEEeqnarray}{RCL}
&&\mathop {\lim }\limits_{{\bm{\vartheta}} \to {{\bm{\theta}}}} {{\bf{B}}^\dag }{\bf{R}}{{\bf{P}}^ \bot } = {\bf{0}} \Rightarrow \mathop {\lim }\limits_{{\bm{\vartheta}} \to {{\bm{\theta}}}} {\bf{\nabla}}  = {\bf{0}},\label{eqn:BRP} \\
&&\mathop {\lim }\limits_{{\bm{\vartheta}} \to {{\bm{\theta}}}} \left(( \ref{eqn:expH1})+(\ref{eqn:expH2}) \right) = ({{\bf{D}}^{\rm H}}{{\bf{P}}^ \bot }{\bf{D}}) \odot {{\bf{\Sigma }}^{\rm T}}.\label{eqn:Hab}
\end{IEEEeqnarray}

According to the theorem on Hadamard product in \cite{Horn1985}, combining Lemma 1 and ${{\bf{\Sigma }}^{\rm T}}\succ0$, we obtain
\begin{equation}
({{{\bf{D}}^{\rm H}}{{\bf{P}}^ \bot }{\bf{D}}) \odot {{\bf{\Sigma}} ^{\rm T}}} \succ 0.
\end{equation}
Note that ${{{\bf{D}}^{\rm H}}{{\bf{P}}^ \bot }{\bf{D}} \odot {{\bf{\Sigma}} ^{\rm T}}}$ is a Hermitian matrix, thus we have
\begin{equation}
\Re (({{\bf{D}}^{\rm{H}}}{{\bf{P}}^ \bot }{\bf{D}}) \odot {{\bf{\Sigma}} ^{\rm{T}}}) \succ 0.
\end{equation}
Combining the results of (\ref{eqn:BRP}) and (\ref{eqn:Hab}), we obtain
\begin{equation}
\mathop {\lim }\limits_{\vartheta  \to {{\bm{\theta}}}} {\bf{H}} = 2\Re \left( {\left( {{{\bf{D}}^{\rm{H}}}{{\bf{P}}^ \bot }{\bf{D}}} \right) \odot {{\bf{\Sigma }}^{\rm{T}}}} \right) \succ 0.
\end{equation}
We thus prove Theorem \ref{Theorem1}.
\section{Proof of Lemma \ref{Lemma2}}\label{ProofLemma2}
In this subsection, ${{\bf{A}}_{-j}}$, ${{\bf{A}}^{-i}}$  and ${A_{i(-j)}}$ are the $j$th to last column, $i$th to last row and $i$th row $j$th to last column of a matrix ${\bf{A}}$, respectively. Let ${\bf{X}} = {{\bf{B}}^\dag }{\bf{A}}{\bf{\Sigma}} {{\bf{A}}^{\rm H}}{{\bf{P}}^ \bot } \odot {{\bf{B}}^{\rm T}}$, we have ${X_{ik}} = {\left( {{{\bf{B}}^\dag }{\bf{A}}{\bf{\Sigma}} {{\bf{A}}^{\rm H}}{{\left( {{{\bf{P}}^ \bot }} \right)}_k}{{\bf{B}}^k}} \right)_{ii}}$.

For arbitrary ULA, whose sensors are located at ${d_1}$, $\cdots$, ${d_M}$, we make a transformation of ${\bf{A}} = {\bf{UG}}$, where ${\bf{G}}$ is a diagonal matrix, whose  $i$th diagonal element is  $\exp \left( {\left( { - j\pi \left( {{d_1} + {d_M}} \right)\sin \left( {{\theta _i}} \right)} \right)/\left( {2\lambda } \right)} \right)$, and ${\bf{U}}$ is an array steering matrix, whose origin is at the middle point of the ULA. Therefore, ${\bf{U}} = {\bf{J}}{{\bf{U}}^*}$, ${\bf{G\Sigma }}{{\bf{G}}^{\rm H}} = {\left( {{\bf{G\Sigma }}{{\bf{G}}^{\rm H}}} \right)^*}$,  where ${\bf{J}}$ is the $M \times M$ exchange matrix whose entries all are zero except the one in the $\left( {i,M - i + 1} \right)$th position for $i = 1, \cdots ,M$. Substituting them into  ${\bf{A\Sigma }}{{\bf{A}}^{\rm H}}$, we have
\begin{align}\label{eqn:ASA}
{\bf{A\Sigma }}{{\bf{A}}^{\rm H}} = {\bf{J}}{\left( {{\bf{A\Sigma }}{{\bf{A}}^{\rm H}}} \right)^*}{\bf{J}}.
\end{align}
Similarly, we let ${\bf{B}} = {\bf{VH}}$ and we have ${\bf{V}} = {\bf{J}}{{\bf{V}}^*}$.
According to the above results and after some matrix manipulations, we obtain
\begin{IEEEeqnarray}{RCL}
{{\bf{B}}^\dag } &=& {{\bf{H}}^{{\rm{ - }}1}}{{\bf{V}}^\dag },\label{eqn:BV}\\
{{\bf{V}}^k} &=& {\left( {{{\bf{V}}^{ - k}}} \right)^*},\label{eqn:Vk}\\
{{\bf{V}}^\dag }{\bf{J}} &=& {\left( {{{\bf{V}}^\dag }} \right)^*},\label{eqn:VPJ}\\
{\bf{J}}{{\bf{P}}^ \bot }_k &=& {\left( {{{\bf{P}}^ \bot }_{ - k}} \right)^*}.\label{eqn:JPO}
\end{IEEEeqnarray}

Substituting (\ref{eqn:BV}) into ${X_{ik}}$ and ${X_{i( - k)}}$, and considering the form of ${\bf{H}}$, we have
\begin{IEEEeqnarray}{RCL}
{X_{ik}} &=& {\left( {{{\bf{V}}^\dag }{\bf{A\Sigma }}{{\bf{A}}^{\rm H}}{{\left( {{{\bf{P}}^ \bot }} \right)}_k}{{\bf{V}}^k}} \right)_{ii}}, \label{eqn:Xik}\\
{X_{i( - k)}} &=& {\left( {{{\bf{V}}^\dag }{\bf{A\Sigma }}{{\bf{A}}^{\rm H}}{{\left( {{{\bf{P}}^ \bot }} \right)}_{ - k}}{{\bf{V}}^{ - k}}} \right)_{ii}}.\label{eqn:XiNk}
\end{IEEEeqnarray}

Substituting (\ref{eqn:ASA})(\ref{eqn:Vk})(\ref{eqn:VPJ})(\ref{eqn:JPO}) into (\ref{eqn:Xik}), we have
\begin{IEEEeqnarray}{RCL}\label{eqn:XikE}
{X_{ik}} &=& \left( {{{\bf{V}}^\dag }{\bf{A\Sigma }}{{\bf{A}}^{\rm{H}}}{{\left( {{{\bf{P}}^ \bot }} \right)}_{ - k}}{{\bf{V}}^{ - k}}} \right)_{ii}^*.
\end{IEEEeqnarray}

Comparing (\ref{eqn:XikE}) with (\ref{eqn:XiNk}), we have ${X_{i\left( { - k} \right)}} = X_{ik}^*$. So we obtain
\begin{align}\label{eqn:Reik}
\Re \left( {{X_{i\left( { - k} \right)}}} \right) = \Re \left( {{X_{ik}}} \right).
\end{align}

Since ${{\bf{P}}^ \bot }{\bf{B}} = {\bf{0}}$, $\sum\nolimits_{k = 1}^M {{{({{\bf{P}}^ \bot })}_k}{B_{ki}}}  = {\bf{0}}$, $\sum\nolimits_{k = 1}^M {{X_{ik}}}  = 0$,
\begin{align}\label{eqn:Reik0}
\sum\nolimits_{k = 1}^M {\Re ({X_{ik}})}  = 0.
\end{align}

Note that ${d_1}$, $\cdots$, ${d_M}$ is a arithmetic progression, and according to (\ref{eqn:Reik}) and (\ref{eqn:Reik0}), we acquire
\begin{IEEEeqnarray}{rCl}\label{eqn:BiA}
&&\Re \left( {\sum\nolimits_{k = 1}^M {{d_k}{X_{ik}}} } \right)=0.
\end{IEEEeqnarray}
Namely, $\sum\nolimits_{k = 1}^M {{d_k}{X_{ik}}} $ is a pure imaginary number.

Since ${D_{ki}} = \alpha_i {d_k}{B_{ki}}$, where $\alpha_i  = {{ - j2\pi \cos \left( {{\vartheta _i}} \right)} \mathord{\left/
 {\vphantom {{ - j2\pi d\cos \left( {{\vartheta _i}} \right)} \lambda }} \right.
 \kern-\nulldelimiterspace} \lambda }$ is a pure imaginary number. And combining (\ref{eqn:BiA}), we have proved that $
{{\bf{B}}^\dag }^i{\bf{A}}{\bf{\Sigma}} {{\bf{A}}^{\rm H}}{{\bf{P}}^ \bot }{{\bf{D}}_i}  = \alpha_i \sum\nolimits_{k = 1}^M {{d_k}{X_{ik}}}$ is real number.
%\begin{IEEEeqnarray}{rCl}
%{{\bf{B}}^\dag }^i{\bf{A}}{\bf{\Sigma}} {{\bf{A}}^{\rm H}}{{\bf{P}}^ \bot }{{\bf{D}}_i} && = \alpha_i \sum\nolimits_{k = 1}^M {{d_k}{X_{ik}}},\IEEEnonumber
%\end{IEEEeqnarray}

\section{Proof of Lemma \ref{Lemma3}}\label{ProofLemma3}
The eigenvalues of ${{\bf{P}}^ \bot }$ are composed of $M - K$ ones and $K$ zeros. Because ${{\bf{P}}^ \bot }{{\bf{e}}_i} = {{\bf{e}}_i}$, ${{\bf{e}}_i}$ is normalized eigenvector of ${{\bf{P}}^ \bot }$ corresponding to eigenvalue 1. Therefore, ${{\bf{P}}^ \bot }$ can be eigendecomposed into ${{\bf{P}}^ \bot }=\sum\nolimits_{i = 1}^{M - K} {{{\bf{q}}_i}{\bf{q}}_i^{\rm H}} $, where ${{\bf{q}}_1} = {{\bf{e}}_i}$. Thus, ${{\bf{P}}^ \bot } - {{\bf{e}}_i}{\bf{e}}_i^{\rm H}{\rm{ = }}\sum\nolimits_{i = 2}^{M - K} {{{\bf{q}}_i}{\bf{q}}_i^{\rm H}} $. So the eigenvalues of ${{\bf{P}}^ \bot } - {{\bf{e}}_i}{\bf{e}}_i^{\rm H}$  are composed of $M - K - 1$  ones and $K + 1$ zeros. Therefore, ${{\bf{P}}^ \bot } - {{\bf{e}}_i}{\bf{e}}_i^{\rm H} \succeq 0$.

\section*{Acknowledgment}

The authors would like to thank...

\ifCLASSOPTIONcaptionsoff
  \newpage
\fi

\bibliographystyle{IEEEtran}
\bibliography{IEEEabrv,ML}

% Generated by IEEEtran.bst, version: 1.13 (2008/09/30)
\begin{thebibliography}{10}
\providecommand{\url}[1]{#1}
\csname url@samestyle\endcsname
\providecommand{\newblock}{\relax}
\providecommand{\bibinfo}[2]{#2}
\providecommand{\BIBentrySTDinterwordspacing}{\spaceskip=0pt\relax}
\providecommand{\BIBentryALTinterwordstretchfactor}{4}
\providecommand{\BIBentryALTinterwordspacing}{\spaceskip=\fontdimen2\font plus
\BIBentryALTinterwordstretchfactor\fontdimen3\font minus
  \fontdimen4\font\relax}
\providecommand{\BIBforeignlanguage}[2]{{%
\expandafter\ifx\csname l@#1\endcsname\relax
\typeout{** WARNING: IEEEtran.bst: No hyphenation pattern has been}%
\typeout{** loaded for the language `#1'. Using the pattern for}%
\typeout{** the default language instead.}%
\else
\language=\csname l@#1\endcsname
\fi
#2}}
\providecommand{\BIBdecl}{\relax}
\BIBdecl

\bibitem{Sandkuhler1987}
U.~Sandkuhler and J.~Bohme, ``Accuracy of maximum-likelihood estimates for
  array processing,'' in \emph{Proc. IEEE Int. Conf. Acoust., Speech, Signal
  Process.}, vol.~12, Apr 1987, pp. 2015--2018.

\bibitem{Stoica1989}
P.~Stoica and N.~Arye, ``Music, maximum likelihood, and cramer-rao bound,''
  \emph{{IEEE} Trans. Acoust., Speech, Signal Process.}, vol.~37, no.~5, pp.
  720--741, May 1989.

\bibitem{Stoica1990}
P.~Stoica and K.~Sharman, ``Maximum likelihood methods for direction-of-arrival
  estimation,'' \emph{{IEEE} Trans. Acoust., Speech, Signal Process.}, vol.~38,
  no.~7, pp. 1132--1143, Jul 1990.

\bibitem{Stoica1990a}
P.~Stoica and A.~Nehorai, ``Music, maximum likelihood, and cramer-rao bound:
  further results and comparisons,'' \emph{{IEEE} Trans. Acoust., Speech,
  Signal Process.}, vol.~38, no.~12, pp. 2140--2150, Dec 1990.

\bibitem{Stoica1990b}
------, ``Performance study of conditional and unconditional
  direction-of-arrival estimation,'' \emph{{IEEE} Trans. Acoust., Speech,
  Signal Process.}, vol.~38, no.~10, pp. 1783--1795, Oct 1990.

\bibitem{Gill1981}
P.~E. Gill, W.~Murray, and M.~H. Wright, \emph{Practical optimization}.\hskip
  1em plus 0.5em minus 0.4em\relax Academic press, 1981.

\bibitem{DennisJr1996}
J.~E. Dennis~Jr and R.~B. Schnabel, \emph{Numerical methods for unconstrained
  optimization and nonlinear equations}.\hskip 1em plus 0.5em minus 0.4em\relax
  Siam, 1996.

\bibitem{Krim1996}
H.~Krim and M.~Viberg, ``Two decades of array signal processing research: the
  parametric approach,'' \emph{{IEEE} Signal Process. Mag.}, vol.~13, no.~4,
  pp. 67--94, Jul 1996.

\bibitem{Schmidt1986}
R.~Schmidt, ``Multiple emitter location and signal parameter estimation,''
  \emph{{IEEE} Trans. Antennas Propag.}, vol.~34, no.~3, pp. 276--280, Mar
  1986.

\bibitem{Donoho2006}
D.~Donoho, ``Compressed sensing,'' \emph{{IEEE} Trans. Inf. Theory}, vol.~52,
  no.~4, pp. 1289--1306, April 2006.

\bibitem{Chen1998}
S.~Chen, D.~Donoho, and M.~Saunders, ``Atomic decomposition by basis pursuit,''
  \emph{SIAM J. Sci. Comput.}, vol.~20, no.~1, pp. 33--61, 1998.

\bibitem{Candes2006}
E.~J. Cand{\`{e}}s, J.~K. Romberg, and T.~Tao, ``Stable signal recovery from
  incomplete and inaccurate measurements,'' \emph{Commun. Pure Appl. Math.},
  vol.~59, no.~8, pp. 1207--1223, 2006.

\bibitem{Li2008}
M.~Li and Y.~Lu, ``Maximum likelihood doa estimation in unknown colored noise
  fields,'' \emph{{IEEE} Trans. Aerosp. Electron. Syst.}, vol.~44, no.~3, pp.
  1079--1090, July 2008.

\bibitem{Mallat1993}
S.~Mallat and Z.~Zhang, ``Matching pursuits with time-frequency dictionaries,''
  \emph{{IEEE} Trans. Signal Process.}, vol.~41, no.~12, pp. 3397--3415, Dec
  1993.

\bibitem{DeVore1996}
R.~DeVore and V.~Temlyakov, ``Some remarks on greedy algorithms,'' \emph{Adv.
  Comput. Math.}, vol.~5, no.~1, pp. 173--187, 1996.

\bibitem{Tipping2001}
M.~E. Tipping, ``Sparse bayesian learning and the relevance vector machine,''
  \emph{J. Mach. Learn. Res.}, vol.~1, pp. 211--244, Sep. 2001.

\bibitem{Ji2008}
S.~Ji, Y.~Xue, and L.~Carin, ``Bayesian compressive sensing,'' \emph{{IEEE}
  Trans. Signal Process.}, vol.~56, no.~6, pp. 2346--2356, June 2008.

\bibitem{Babacan2010}
S.~Babacan, R.~Molina, and A.~Katsaggelos, ``Bayesian compressive sensing using
  laplace priors,'' \emph{{IEEE} Trans. Image Process.}, vol.~19, no.~1, pp.
  53--63, Jan 2010.

\bibitem{Malioutov2005}
D.~Malioutov, M.~Cetin, and A.~Willsky, ``A sparse signal reconstruction
  perspective for source localization with sensor arrays,'' \emph{{IEEE} Trans.
  Signal Process.}, vol.~53, no.~8, pp. 3010--3022, Aug 2005.

\bibitem{Cotter2005}
S.~Cotter, B.~Rao, K.~Engan, and K.~Kreutz-Delgado, ``Sparse solutions to
  linear inverse problems with multiple measurement vectors,'' \emph{{IEEE}
  Trans. Signal Process.}, vol.~53, no.~7, pp. 2477--2488, July 2005.

\bibitem{Eldar2009}
Y.~Eldar and M.~Mishali, ``Robust recovery of signals from a structured union
  of subspaces,'' \emph{{IEEE} Trans. Inf. Theory}, vol.~55, no.~11, pp.
  5302--5316, Nov 2009.

\bibitem{Tropp2006}
J.~A. Tropp, ``Algorithms for simultaneous sparse approximation. {Part II}:
  Convex relaxation,'' \emph{Signal Process.}, vol.~86, no.~3, pp. 589 -- 602,
  2006.

\bibitem{Hyder2009}
M.~Hyder and K.~Mahata, ``A robust algorithm for joint-sparse recovery,''
  \emph{{IEEE} Signal Process. Lett.}, vol.~16, no.~12, pp. 1091--1094, Dec
  2009.

\bibitem{Tropp2006a}
J.~A. Tropp, A.~C. Gilbert, and M.~J. Strauss, ``Algorithms for simultaneous
  sparse approximation. {Part I}: Greedy pursuit,'' \emph{Signal Process.},
  vol.~86, no.~3, pp. 572 -- 588, 2006.

\bibitem{Wipf2007}
D.~Wipf and B.~Rao, ``An empirical bayesian strategy for solving the
  simultaneous sparse approximation problem,'' \emph{{IEEE} Trans. Signal
  Process.}, vol.~55, no.~7, pp. 3704--3716, July 2007.

\bibitem{Zhu2011}
H.~Zhu, G.~Leus, and G.~Giannakis, ``Sparsity-cognizant total least-squares for
  perturbed compressive sampling,'' \emph{{IEEE} Trans. Signal Process.},
  vol.~59, no.~5, pp. 2002--2016, May 2011.

\bibitem{Yang2013}
Z.~Yang, L.~Xie, and C.~Zhang, ``Off-grid direction of arrival estimation using
  sparse bayesian inference,'' \emph{{IEEE} Trans. Signal Process.}, vol.~61,
  no.~1, pp. 38--43, Jan 2013.

\bibitem{Pesavento2001}
M.~Pesavento and A.~Gershman, ``Maximum-likelihood direction-of-arrival
  estimation in the presence of unknown nonuniform noise,'' \emph{{IEEE} Trans.
  Signal Process.}, vol.~49, no.~7, pp. 1310--1324, Jul 2001.

\bibitem{Ottersten1993}
B.~Ottersten, M.~Viberg, P.~Stoica, and A.~Nehorai, in \emph{Radar Array
  Processing}.\hskip 1em plus 0.5em minus 0.4em\relax Springer Berlin
  Heidelberg, 1993, pp. 99--151.

\bibitem{Golub1996}
G.~H. Golub and C.~F. van Van~Loan, \emph{Matrix computations}.\hskip 1em plus
  0.5em minus 0.4em\relax The Johns Hopkins University Press, 1996.

\bibitem{Golub1973}
G.~Golub and V.~Pereyra, ``The differentiation of pseudo-inverses and nonlinear
  least squares problems whose variables separate,'' \emph{SIAM J. Numer.
  Anal.}, vol.~10, no.~2, pp. 413--432, 1973.

\bibitem{Viberg1991}
M.~Viberg and B.~Ottersten, ``Sensor array processing based on subspace
  fitting,'' \emph{{IEEE} Trans. Signal Process.}, vol.~39, no.~5, pp.
  1110--1121, May 1991.

\bibitem{Horn1985}
R.~A. Horn and C.~R. Johnson, \emph{Matrix Analysis}.\hskip 1em plus 0.5em
  minus 0.4em\relax Cambridge University Press, 1985.

\end{thebibliography}
%\bibliography{ML}

\end{document}